\documentclass[sigconf]{acmart}
\usepackage{float}
\usepackage{xcolor}

\definecolor{revcolor}{HTML}{0B5FA5}

\newcommand\blfootnote[1]{%
  \begingroup
  \renewcommand\thefootnote{}\footnote{#1}%
  \addtocounter{footnote}{-1}%
  \endgroup
}

\AtBeginDocument{%
  }

\setcopyright{acmlicensed}
\copyrightyear{2026}
\acmYear{2026}

\acmConference[UMFM '26]
{The 2nd ACM SIGSPATIAL International Workshop on Urban Mobility Foundation Models}
{November 3, 2026}
{Riverside, CA, USA}

\acmBooktitle{UMFM '26: The 2nd ACM SIGSPATIAL International Workshop on Urban Mobility Foundation Models, November 3, 2026, Riverside, CA, USA}
\begin{document}

\title{A Graph Neural Network Surrogate Model for Incident-Based Travel Time Prediction Under Limited Sensor Data}

\author{Abhilasha Saroj}
\authornote{Corresponding author: Abhilasha Saroj (sarojaj@ornl.gov).}
\authornote{These authors contributed equally to this work and share first authorship.}
\affiliation{%
  \institution{Oak Ridge National Laboratory}
  \city{Oak Ridge}
  \state{Tennessee}
  \country{USA}
}

\author{Natalie Myers}
\authornotemark[2]
\affiliation{%
  \institution{Oak Ridge National Laboratory}
  \city{Oak Ridge}
  \state{Tennessee}
  \country{USA}
}

\author{Haoran Niu}
\affiliation{%
  \institution{Oak Ridge National Laboratory}
  \city{Oak Ridge}
  \state{Tennessee}
  \country{USA}
}

\renewcommand{\shortauthors}{ Saroj, Myers, Niu}

\begin{abstract}
Localized roadway incidents produce congestion far beyond their point of origin, but evaluating this with microscopic simulation requires a full run per scenario. This paper presents a Spatio-Temporal Graph Convolutional Network (STGCN) that forecasts route-level travel times on a simulated Nashville, Tennessee area road network consisting of 1,037 junctions and 1,601 road segments, where directional traffic counts at its 129 signalized intersections are the only traffic input. That restriction is deliberate: signalized intersections are where sensing already exists, so the input design targets data agencies already collect. By representing those intersections as a graph, the STGCN captures spatial dependence among connected locations and the temporal evolution of traffic jointly, learning how localized disturbances propagate. Two models are compared: a baseline trained on 80 incident-free simulations, and an incident-inclusive model adding 360 lane-blockage scenarios across 12 locations and 3 durations. The baseline achieved a test mean absolute error (MAE) of 109.68 seconds, a relative MAE of 9.94\%, that is, total absolute forecast error as a share of total observed travel time across the twelve monitored routes. The incident-inclusive model achieved a relative MAE of 9.67\% on the full test set, 6.45\% during active incidents, and 13.62\% at thirty incident locations withheld entirely from training, and on the disrupted route its forecasts fall within 2.0 minutes of observed travel time on average. Travel times in the simulations themselves vary by 7.4\%, or 1.3 minutes, between random seeds under identical conditions, so prediction error is of the same order as the network's own stochastic variability. Producing a forecast takes roughly 53~milliseconds on a single processor core, making the model usable as a computationally efficient surrogate for transportation resilience screening, incident management, and proactive re-routing.
\end{abstract}



\maketitle

\blfootnote{This manuscript has been authored in part by UT-Battelle, LLC, under contract DE-AC05-00OR22725 with the U.S. Department of Energy (DOE). The U.S. Government retains and the publisher, by accepting the article for publication, acknowledges that the U.S. Government retains a non-exclusive, paid-up, irrevocable, worldwide license to publish or reproduce the published form of this manuscript, or allow others to do so, for U.S. Government purposes. DOE will provide public access to these results of federally sponsored research in accordance with the DOE Public Access Plan (https://energy.gov/doe-public-access-plan).}

\section{Introduction}

Urban traffic congestion imposes substantial economic and societal costs, and localized disruptions such as lane blockages can propagate well beyond their point of origin, producing congestion across large portions of a network. Forecasting these disturbances is important for traffic management and incident response, but high-fidelity microscopic simulators, which model every vehicle individually, require a complete simulation for every scenario evaluated, making large-scale incident analysis computationally expensive.

Graph neural networks (GNNs) suit this problem because roadway networks are inherently graph-structured, with junctions as nodes and road segments as the links between them, and because repeatedly combining each node's state with those of the nodes it connects to is a direct analogue of a queue at one junction spilling back into the ones feeding it. Spatio-Temporal Graph Convolutional Networks (STGCNs) add temporal convolution to this, learning how congestion develops locally before propagating through adjacent corridors and capturing spatial and temporal dependence jointly.

This paper develops such a model for Nashville, Tennessee, forecasting route-level travel times from directional traffic counts recorded at signalized intersections, together with the pipeline that extracts that data from simulation and assembles it as graph-structured input. The testbed is a microscopic simulation of a real urban network, driven by field-derived traffic demand and signal control. Two models are trained on its output: a baseline model seeing only ordinary, incident-free traffic, and an incident-inclusive model trained on that same data together with simulated lane blockages, which is further tested on incident locations withheld entirely from training.

This paper makes two key contributions. First, it shows that route-level travel time can be forecast from directional traffic counts measured only at signalized intersections, the subset of a network where near-real-time data is routinely available, rather than from dense sensing along the routes being predicted. Second, and most relevant to reusable mobility models, it evaluates whether incident response transfers to disruption locations withheld entirely from training, a question the prior work this architecture builds on does not address, reporting both the degree of transfer and where it degrades.



\subsection{Terminology}
\label{sec:terminology}

Several terms carry a specific meaning here and are used consistently throughout. A \emph{junction} is a point where roads meet and corresponds to a graph node. Of the 1,037 junctions in the study network, the 129 \emph{signalized intersections} are the only ones supplying model input, while \emph{unsignalized} junctions are retained for connectivity but carry no measurement. A \emph{road segment} is a directed stretch of road between two junctions, identified in figures and tables by its Simulation of Urban MObility (SUMO)~\cite{lopez2018sumo} edge identifier. A \emph{route} is a fixed, ordered sequence of segments forming one trip, and twelve routes are monitored as prediction targets.

A \emph{blockage} is a lane closed to traffic, reproduced in simulation by stopping a vehicle in the travel lane so that traffic must merge around it and a queue forms behind. The term \emph{incident} is used interchangeably for the resulting event and its network-wide effects. \emph{Baseline} denotes an otherwise identical run with no blockage inserted. Because a baseline run and its incident counterpart share a random \emph{seed}, and so generate the same vehicles making the same trips, the difference between the pair isolates the delay the blockage caused. \emph{Directional traffic count}, the model's input, is the vehicle count on segments entering a signalized intersection, averaged over one minute and grouped into northbound, southbound, eastbound, and westbound. The \emph{observation window} is the span of past minutes read as input, and the \emph{forecast horizon} is the span of future minutes predicted.

\section{Background and Motivation}
\label{sec:background}

Operational traffic data is collected at points, not continuously. Sensing concentrates at signalized intersections, where inductive loops, radar, and video analytics report approach-level counts and occupancy at sub-minute resolution because signal control requires them. The segments between are typically unsensed. Route travel time, the quantity operators and travelers actually act on, is therefore not measured directly across most of a network. Probe-vehicle feeds give partial coverage but are spatially aggregated, depend on how many vehicles participate, and lag the onset of a disruption, which is exactly when the information matters most.

This paper asks whether route-level travel time can be forecast a short horizon ahead from that sparse intersection data alone, under both recurring conditions and active incidents. A graph formulation fits the problem: the instrumented points are junctions in a physical network, and the mechanism linking a local disruption to a distant delay, queue formation and spillback, is propagation along that network. Hence the design choice carried throughout this work, that the model consumes directional traffic counts only at the 129 signalized intersections, while the remaining 908 junctions are retained solely to preserve topology. No measurement is assumed that an agency would not already have.

Training such a model on field data is difficult because the counterfactual is never observed: for a real incident there is no measurement of what travel times would have been without it, which is what isolates incident-induced delay. Microscopic simulation supplies it directly, since paired runs sharing a random seed can differ only in the presence of a blockage, and location, duration, and demand can be varied systematically. The practical barrier is the cost of building and calibrating network models, which automated scenario generation~\cite{xu2025realtwin} and digital twin research~\cite{saroj2021dt, saroj2023imputation, roy2024evp} address. This work is complementary, treating the simulation as a data generator for a learned surrogate that could later be re-grounded on a field-calibrated twin without changing its inputs.

The motivation for a surrogate is throughput: simulation must run to completion for every scenario, while operationally useful analyses are combinatorial across locations, durations, demand levels, and seeds. The trained model forecasts over the full 1,037-junction graph in roughly 53~milliseconds on a single processor core, with about 55,000 parameters. That supports rapid resilience screening across many candidate locations, incident-management estimates produced while a blockage is still active, and route choice informed by forecast rather than current conditions. Screening in particular depends on transferability rather than accuracy alone, since the locations of interest are by definition ones not yet simulated. Section~\ref{sec:transfer} evaluates this directly.


\section{Related Work}

Deep spatio-temporal models have become the dominant approach to traffic forecasting, representing the road network as a graph and pairing graph convolution with temporal modeling. Yu et al.~\cite{yu2018stgcn} introduced the STGCN, which replaces recurrent temporal modeling with gated temporal convolution and Chebyshev graph convolution inside a repeatable ST-Conv block, training faster and with fewer parameters than recurrent alternatives while matching or exceeding their accuracy; that block is the architectural basis used here. The block is reused, but the task differs: Yu et al. forecast the same quantity at the same locations that supply the input, whereas this model reads approach counts at 129 signalized intersections and predicts travel time on twelve routes. The target is also harder in kind, being the travel time of a vehicle departing in a given minute rather than the network state at that minute. Later variants add diffusion-based directional flow~\cite{li2018dcrnn}, spatial and temporal attention~\cite{guo2019astgcn}, learned rather than topological adjacency~\cite{wu2019graphwavenet}, node-specific parameters~\cite{bai2020agcrn}, and lightweight recurrent-convolutional hybrids~\cite{zhao2020tgcn}. Surveys~\cite{jiang2022survey} report a consistent evaluation setting across this family: loop-detector archives such as METR-LA and PEMS, dense freeway sensing, prediction of the measured quantity at the measured locations, and overwhelmingly recurring conditions. This paper departs on all three counts, predicting route travel time from intersection counts, sensing 129 of 1,037 junctions, and targeting non-recurring conditions. Those departures also explain the absence of a benchmark comparison: no public dataset known to us supplies paired incident and incident-free observations of one network under identical demand, which the counterfactual here requires.

Fan et al.~\cite{fan2026igstgnn} show that training on mixed normal and incident data without incident labels yields an averaged function that smooths over the sharp deviations incidents cause, and address this architecturally with Incident-Guided Spatiotemporal Graph Neural Network (IGSTGNN), which fuses incident context by attention and decays its influence across the horizon. This paper targets the same limitation through the training data instead, admitting incidents as a binary feature channel and up-weighting incident samples in the loss. Neither work tests whether incident response learned at fixed locations transfers to locations excluded from training, which is evaluated here. A second line of work approaches the problem from the simulation side: microscopic simulators reproduce incident effects mechanistically at high cost per scenario, and digital twin research keeps such models aligned with field conditions, spanning real-time corridor simulation~\cite{saroj2021dt}, imperfect and imputed sensor inputs~\cite{saroj2023imputation}, automated scenario generation~\cite{xu2025realtwin}, and operational uses such as emergency vehicle preemption~\cite{roy2024evp}. The contribution here is to use that representation as a training source for a learned surrogate, retaining the simulator for scenarios that warrant full fidelity.

\section{Data and Methodology}

\subsection{SUMO Traffic Simulation Model as Graph}

Traffic simulations were conducted in SUMO and controlled through the Traffic Control Interface (TraCI), enabling automated scenario generation and insertion or removal of lane-blocking incidents without modifying network or demand files.

\begin{figure}[tb]
  \centering
  \includegraphics[width=\columnwidth]{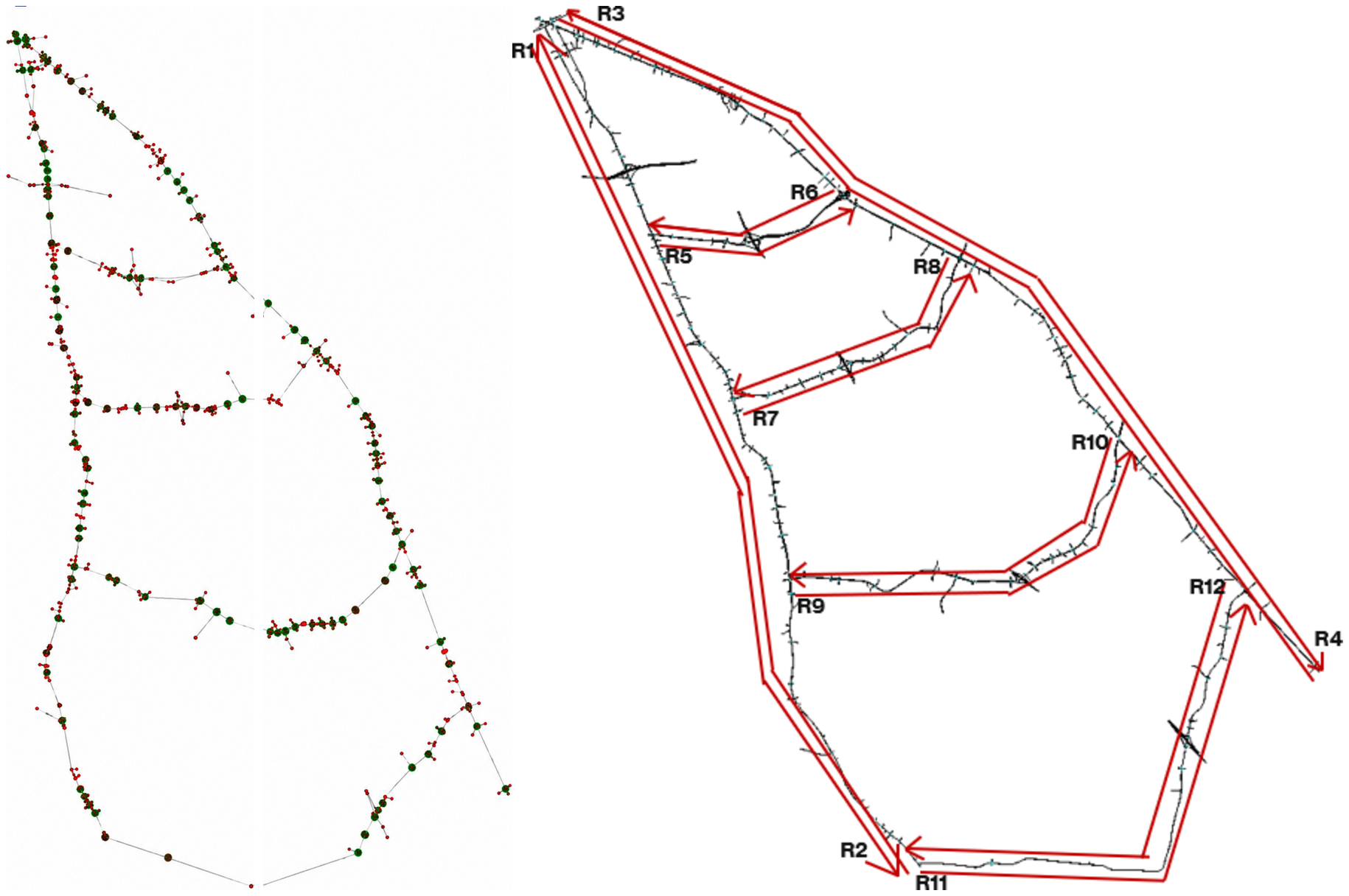}
  \caption{Left: signalized (green) and unsignalized (red) junctions across the study network. Right: the twelve monitored routes (R1--R12) overlaid on the same network.}
  \Description{Two-panel figure. Left panel shows the road network with signalized intersections marked in green and unsignalized intersections marked in red. Right panel shows the same network with the twelve prediction-target routes traced and labeled R1 through R12.}
  \label{fig:network_routes}
\end{figure}

The Nashville network contains 1,037 junctions, 1,601 directed road segments, and 129 signalized intersections. The network was converted into a directed graph using \texttt{sumolib} and NetworkX. Twelve routes were selected as prediction targets: four follow the primary commuter corridors, Nolensville Pike and Murfreesboro Pike, with free-flow travel times of approximately 20 to 30 minutes, while eight shorter routes connect these corridors to provide broad spatial coverage. Figure~\ref{fig:network_routes} shows these routes together with the distribution of signalized and unsignalized junctions across the network.

Route travel times were measured using dedicated ego vehicles (probe vehicles inserted solely for measurement), as SUMO's E3 detectors report averaged statistics rather than complete end-to-end travel times. One ego vehicle was inserted onto each route every minute, with travel time computed from insertion to arrival. Each measurement is recorded against the one-minute bin in which the vehicle departed, so the prediction target is the travel time experienced by a vehicle departing in that minute rather than a snapshot of network conditions at that minute. This distinction matters for interpreting the forecast horizon: on the longest routes, a departure occurring at the end of the horizon does not complete its trip until roughly half an hour later, so the model is implicitly anticipating conditions beyond the horizon itself. This is the quantity of operational interest, since it is what a traveler departing at a given minute would experience, but it is a stronger requirement than forecasting instantaneous network state. Network-wide road segment measurements, including vehicle count, speed, travel time, and waiting time, were collected every minute across the entire network so that directional traffic counts reflected congestion propagation beyond the monitored routes. To avoid unrealistically large travel-time estimates during severe congestion, segments with mean speed below 1 m/s were assigned free-flow travel time plus mean waiting time.

\begin{figure}[tb]
  \centering
  \includegraphics[width=\columnwidth]{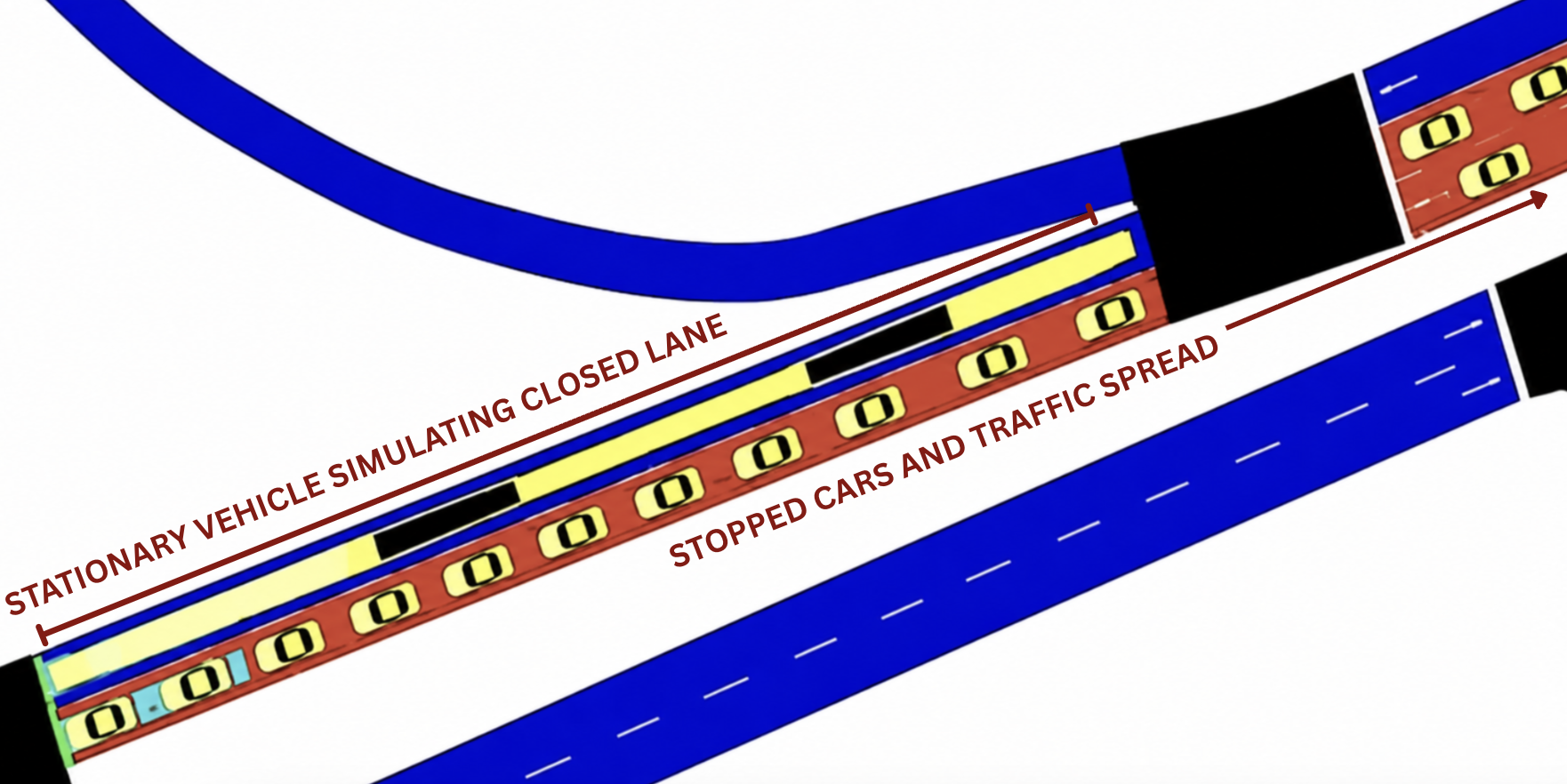}
  \caption{Example of a simulated lane-blockage incident in SUMO, showing a stationary vehicle obstructing one lane while adjacent traffic continues to flow.}
  \Description{Screenshot of the SUMO simulator showing a stationary blocking vehicle occupying one lane of a road segment, with vehicle icons queuing behind it.}
  \label{fig:blockage_example}
\end{figure}

Incidents were modeled as stationary vehicles blocking a lane, producing realistic upstream queuing while maintaining valid routes (Figure~\ref{fig:blockage_example}).

Each simulation covered 8:00 to 10:00 a.m., with incidents beginning at 8:15 a.m. and lasting 30, 40, or 50 minutes. Twelve incident locations (identified in figures and tables by their SUMO edge identifiers) and 30 random seeds per location-duration combination produced 1,080 incident scenarios, together with 80 incident-free baseline simulations. Paired baseline and incident runs shared random seeds to isolate incident effects. Driver heterogeneity was represented by assigning 20\% of demand more aggressive and 20\% more cautious lane-changing behavior.

For transferability evaluation, an additional thirty incident locations, distinct from the twelve used in training, were simulated at a fixed 40-minute duration across five random seeds each.

Each minute, incoming road segments at every signalized intersection were classified into northbound, southbound, eastbound, or westbound approaches, producing four directional count channels per intersection. Unsignalized junctions remained in the graph to preserve network topology. Because features are supplied only at the 129 signalized intersections, the remaining 908 junctions carry zero-valued feature vectors and contribute to prediction only through the graph convolution, which is what allows the input specification to match the sensing an agency would realistically have available. A fifth feature channel indicated incident status, remaining 0 during baseline conditions and 1 throughout active incidents, reflecting the assumption that incidents have already been detected by the time a forecast is requested.

Training samples used a 15-minute sliding observation window paired with the 15 minutes of travel time immediately following it. Inputs have dimensions $[15, 1037, 5]$ and stored targets have dimensions $[15, 12]$, that is, 15 minutes for each of the 12 monitored routes. Because each ST-Conv block shortens the sequence by $2(k-1)$ steps, the model returns 11 of these 15 steps, and the loss is computed against the final 11 target minutes. The forecast horizon is therefore minutes 5 through 15 following the end of the observation window, rather than the first 11 minutes after it. Data were divided 70/15/15 into training, validation, and test sets by random assignment of sliding windows, stratified by incident status. Because consecutive windows from one run overlap by fourteen of their fifteen minutes, windows closely resembling test samples also appear in training, so the held-out figures reported here are optimistic relative to a split by simulation run or by random seed.

\subsection{Model Architecture and Training}


\begin{figure*}[t]
  \centering
  \includegraphics[width=\textwidth]{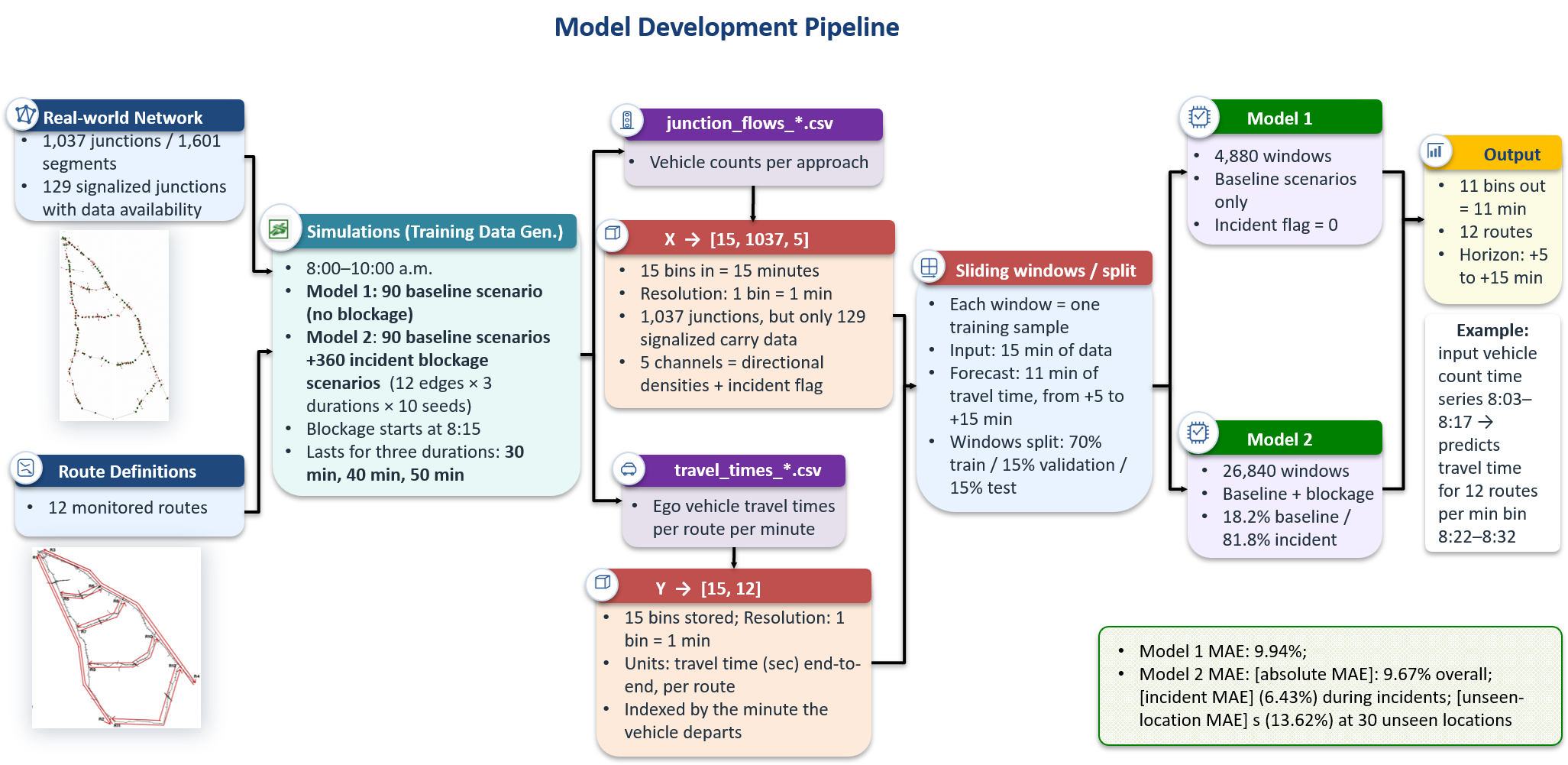}
  \caption{Overview of the full pipeline, from SUMO simulation and data extraction through feature construction, the STGCN model, and evaluation.}
  \Description{Flowchart showing the process from SUMO simulation through TraCI data extraction, feature construction, the STGCN model, route-level predictions, and the three evaluation stages.}
  \label{fig:pipeline}
\end{figure*}

Figure~\ref{fig:pipeline} summarizes the full pipeline, from simulation and data extraction through feature construction to the model and its evaluation. Traffic forecasting was formulated as a spatio-temporal graph learning problem using the STConv layer from PyTorch Geometric Temporal. Each block consists of a temporal convolution, a Chebyshev graph convolution, and a second temporal convolution, jointly capturing temporal evolution and spatial congestion propagation:

\begin{equation}
\Gamma(X)=\mathrm{ReLU}\!\left(\left(X*W_1\right)\odot\sigma\!\left(X*W_2\right)+\left(X*W_3\right)\right)
\end{equation}
\begin{equation}
\Theta *_G X= \sum_{k=0}^{K-1}\theta_k T_k(\tilde{L})X
\end{equation}
\begin{equation}
H^{(l+1)} = \mathrm{BN}\left(\Gamma_2\left(\mathrm{ReLU}\left(\Theta *_G\left(\Gamma_1(H^{(l)})\right)\right)\right)\right)
\end{equation}

In the temporal gated convolution (Eq.~1), $X$ is the input sequence for a given node. $W_1$, $W_2$, and $W_3$ are three learnable convolutional kernels applied to $X$, producing a linear output, a gating signal, and a residual term. The function $\sigma(\cdot)$ is the sigmoid, which maps the gating signal to a value between 0 and 1, and $\odot$ denotes element-wise multiplication. $\mathrm{ReLU}(\cdot)$ is the rectified linear unit. The linear output $X*W_1$ is scaled element-wise by the gate $\sigma(X*W_2)$, so that the gate controls how much of each time step's information passes forward, and the residual term $X*W_3$ is added to this gated signal before the nonlinearity, allowing the block to preserve information that the gate would otherwise suppress. This gated convolutional structure lets the model down-weight less relevant time steps within the input window rather than treating every observed minute equally. Because each temporal convolution consumes $k-1$ time steps and an ST-Conv block applies two of them, an input sequence of length $m$ produces an output of length $m-2(k-1)$. With $m=15$ and kernel width $k=3$, this yields the 11-step output used throughout this paper.

In the graph convolution (Eq.~2), $\Theta$ is the learnable filter, parameterized by coefficients $\theta_k$ for $k = 0, \ldots, K-1$. The Chebyshev order $K$ controls how many hops away from a given node the convolution can draw information from in a single layer. $T_k(\tilde{L})$ is the Chebyshev polynomial of order $k$ evaluated at $\tilde{L}$, the scaled and normalized graph Laplacian, which encodes the network's junction-to-junction connectivity, and $X$ here is the node feature matrix being convolved. This formulation approximates full spectral graph convolution without directly computing the graph's eigendecomposition, which would be too computationally expensive for a network of this size.

Equation~3 combines both operations into a single ST-Conv block: $H^{(l)}$ is the hidden representation entering block $l$, $\Gamma_1$ and $\Gamma_2$ are the first and second temporal gated convolutions within the block, and $\Theta *_G$ is the graph convolution applied between them, followed by a rectified linear unit. $\mathrm{BN}$ denotes batch normalization applied over nodes at the block output. The temporal convolution learns short-term patterns within the input sequence, while the Chebyshev graph convolution aggregates information from neighboring nodes up to $K-1$ hops away to represent congestion propagation through the network. These operations combine within each ST-Conv block to capture both the evolution and spatial distribution of traffic conditions.

The roadway graph used a fixed directed adjacency matrix with unit edge weights, since STConv accepts only scalar edge weights. Binary connectivity was retained to preserve physical roadway structure. A linear layer maps the output channels of the ST-Conv block to a single value at every junction and timestep, giving a scalar field of shape $[\text{batch}, 11, 1037]$, and the prediction for a route at a given timestep is the unweighted mean of that field over the junctions the route traverses. All twelve route outputs are therefore read from one shared node field rather than from separate per-route heads. Inputs and targets enter the model in their raw units, vehicles per minute and seconds respectively, with no feature scaling or target normalization applied at any stage. A single ST-Conv block was used. The resulting model has approximately 55,000 learned parameters, which is small by the standards of this literature and is what makes the inference cost reported in Section~\ref{sec:background} achievable on commodity hardware.

Both models were trained with the Adam optimizer using mean absolute error (MAE) loss. Hyperparameters for each model were tuned independently on the validation set, with the test set held unseen throughout tuning, searching hidden/output channels (32, 64, 128), learning rate (0.005--0.02), weight decay (0, $10^{-4}$), batch size (16, 32), kernel width (1, 2, 3), and Chebyshev order $K \in \{1,2,3\}$.

\subsection{Model 1}

\subsubsection{Architecture}
The baseline model used 32 hidden/output channels, kernel width 3, Chebyshev order $K=2$, and learning rate 0.005, selected from the hyperparameter search described above. Training ran for 50 epochs, with the best validation checkpoint selected by lowest validation MAE.

Training was conducted on an NVIDIA T1000 (8 GB) graphics processing unit (GPU) using PyTorch 2.5.1 (CUDA 12.1) and PyTorch Geometric Temporal, on a Dell Precision 3660 running Python 3.12.

\subsubsection{Training Data}
The dataset comprised the 80 incident-free baseline simulations, representing recurring congestion under normal operating conditions with no incident signal present.

\subsubsection{Testing Data}
Evaluation used the 15\% test split of the 80 baseline simulations. Two complementary quantities are reported. Aggregate accuracy is reported as MAE in seconds and as \emph{relative MAE}, the summed per-route MAE divided by the summed mean observed travel time,
\begin{equation}
\label{eq:relmae}
\text{relative MAE}=\frac{\sum_{r=1}^{R}\mathrm{MAE}_r}{\sum_{r=1}^{R}\bar{y}_r}\times100\%,
\end{equation}
where $\mathrm{MAE}_r$ is the mean of $|\hat{y}-y|$ over every test sample and every forecast timestep for route $r$, and $\bar{y}_r$ is the mean observed travel time over those same elements. Relative MAE reads as the percentage of a trip's duration by which a single forecast is typically wrong. Two properties matter for interpreting it: it is built from absolute errors, so over- and under-predictions never cancel, and it is weighted by travel time, so the long commuter corridors contribute proportionally more than the short connectors, making it a network-level figure rather than an unweighted average over routes. Absolute error is used for these aggregate figures rather than signed delta, to avoid overstating accuracy through cancellation between over- and under-predictions of similar magnitude. Per-route results are additionally reported as the difference between mean predicted and mean actual travel time, which measures systematic bias, that is, whether the model tends to over- or under-predict a given route, rather than the size of a typical individual error. The two are reported side by side because they answer different questions, and they are clearly distinguished wherever both appear.


\subsection{Model 2}

\subsubsection{Architecture}
The incident-inclusive model used 64 hidden/output channels, kernel width 3, and Chebyshev order $K=3$, trained for 200 epochs with learning rate 0.008, weight decay $10^{-4}$, and batch size 32. 
Training was conducted on the CADES/Pathfinder high-performance computing cluster at Oak Ridge National Laboratory, using NVIDIA A2 GPU nodes under SLURM job scheduling, with the same software stack as Model 1.

\subsubsection{Training Data}
The dataset comprised the same 80 baseline simulations used for Model 1, combined with 360 incident scenarios (12 blockage locations $\times$ 3 durations $\times$ 10 random seeds), allowing the network to learn both recurring congestion and incident-driven disruption.

\subsubsection{Testing Data}
\label{sec:model2test}
Three test evaluations were performed. First, the full 15\% held-out test split (baseline and incident samples combined) was evaluated to measure overall forecasting accuracy. Second, evaluation was restricted to incident-affected test samples specifically, timesteps during which an active blockage was present, to isolate incident-response accuracy from the larger, easier baseline population. Third, the model was evaluated on thirty incident locations excluded entirely from training, at a fixed 40-minute duration, to test whether learned incident-response behavior generalizes to previously unseen roadway locations.
\section{Experiments and Results}

\subsection{Model 1: Baseline Performance}

\begin{figure*}[t]
  \centering
  \includegraphics[width=\textwidth]{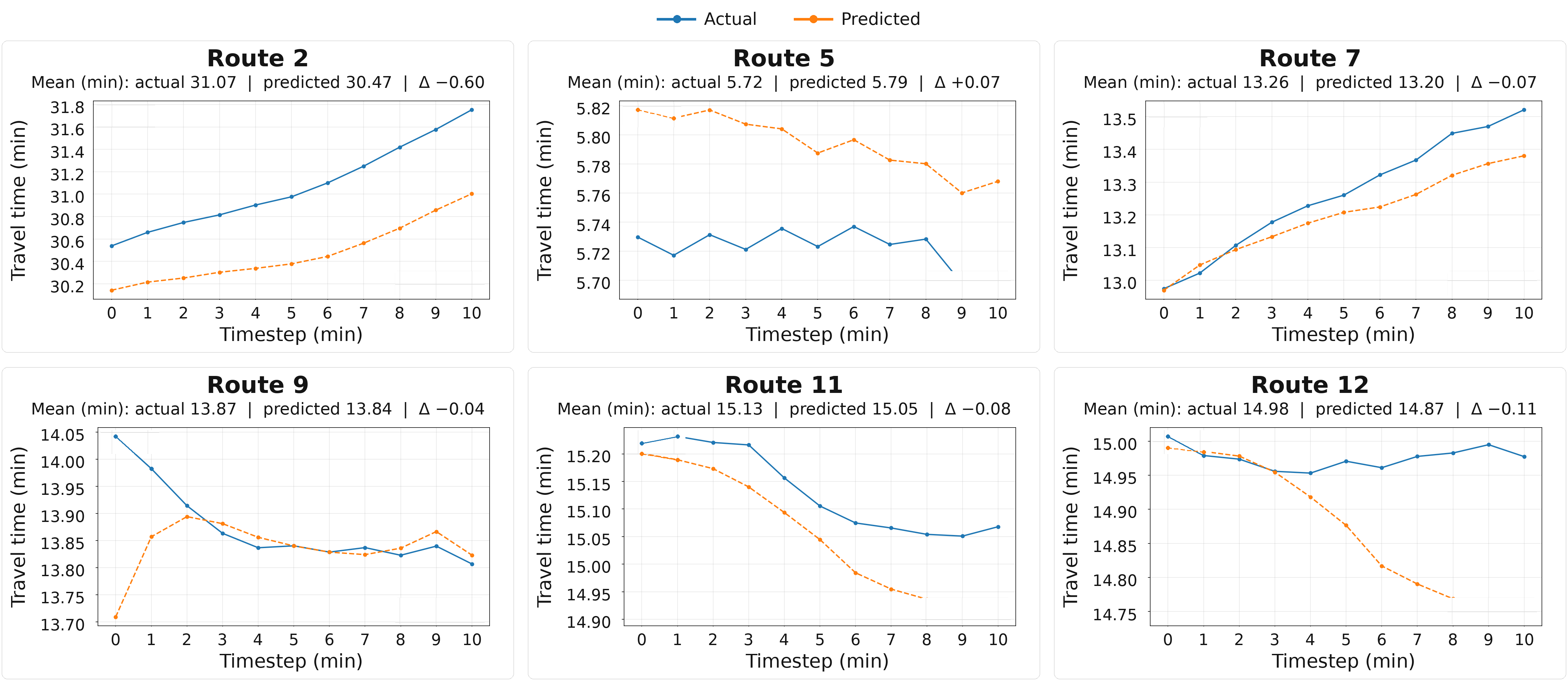}
  \caption{Predicted versus actual travel time curves for six representative routes under Model 1. Y-axis ranges are scaled to each route's own travel-time magnitude to make the small prediction-to-actual deltas visible. At a shared, wider scale these differences would appear visually indistinguishable.}
  \Description{Grid of six line plots, one per route, each showing predicted and actual travel time over the forecast window, with y-axis scaling adjusted per route to reveal small differences between the two lines.}
  \label{fig:model1_curves}
\end{figure*}

The baseline model achieved a best validation MAE of 77.42 seconds. Figure~\ref{fig:model1_curves} shows predicted and actual travel-time curves for six representative routes. Evaluation on the held-out test set produced an overall MAE of 109.68 seconds against a mean observed travel time of 1103.22 seconds across the twelve monitored routes, a relative MAE of 9.94\% by Eq.~\ref{eq:relmae}.


Predicted travel times closely matched observed values for eleven of the twelve routes, with relative differences generally below 2\% and travel-time differences typically within 30 seconds (Table~\ref{tab:baseline_routes}). Route 10 was the only exception, with a $-166.68$ second difference ($-10.68\%$), substantially larger than any other route, indicating that systematic bias was concentrated on a single corridor rather than distributed across the network.

These per-route figures are considerably smaller than the 109.68-second MAE reported above, and the two should not be read as competing estimates of the same quantity. Averaging the twelve absolute per-route differences gives 25.3 seconds, roughly a quarter of the MAE, because the per-route figures compare mean predicted against mean actual travel time and therefore allow over- and under-predictions on individual samples to cancel. The MAE retains those errors, so it is the appropriate measure of how far a single forecast can be expected to fall from the truth, while Table~\ref{tab:baseline_routes} shows that, on eleven of twelve routes, those errors are close to unbiased.


An MAE of 109.68 seconds is difficult to judge in isolation, so it is useful to establish how much travel time varies in this network for reasons no model can predict. Across the 80 incident-free simulations, which differ only in random seed, travel time on the same route at the same clock minute varies with a mean absolute deviation of 79.1 seconds, or 7.4\% of mean route travel time, about 72\% of the model's 109.68-second MAE. Prediction error is therefore of the same order as the variation the network exhibits for reasons unrelated to any model. That comparison is a reference scale rather than a decomposition or a lower bound: the two statistics are constructed differently, and the model additionally observes the recent history of the run it is forecasting.

\begin{table}[tb]
  \caption{Model 1 (baseline) per-route test performance. Actual and Predicted are means over the test split, so Delta and Rel. diff. measure systematic bias, not typical individual error.}
  \label{tab:baseline_routes}
  \centering
  \begin{tabular}{lcccc}
    \toprule
    Route & Actual (s) & Predicted (s) & Delta (s) & Rel. diff. \\
    \midrule
    1  & 1550.97 & 1521.57 & $-29.40$  & $-1.90\%$ \\
    2  & 1862.09 & 1827.00 & $-35.10$  & $-1.88\%$ \\
    3  & 1822.82 & 1844.61 & $21.79$   & $1.20\%$ \\
    4  & 1730.35 & 1746.21 & $15.86$   & $0.92\%$ \\
    5  & 343.34  & 347.73  & $4.38$    & $1.28\%$ \\
    6  & 361.40  & 352.80  & $-8.60$   & $-2.38\%$ \\
    7  & 794.90  & 791.19  & $-3.71$   & $-0.47\%$ \\
    8  & 572.12  & 576.65  & $4.54$    & $0.79\%$ \\
    9  & 832.75  & 830.21  & $-2.53$   & $-0.30\%$ \\
    10 & 1561.05 & 1394.37 & $-166.68$ & $-10.68\%$ \\
    11 & 908.32  & 903.48  & $-4.86$   & $-0.53\%$ \\
    12 & 898.55  & 892.68  & $-5.86$   & $-0.65\%$ \\
    \bottomrule
  \end{tabular}
\end{table}

\begin{figure}[tb]
  \centering
  \includegraphics[width=\columnwidth]{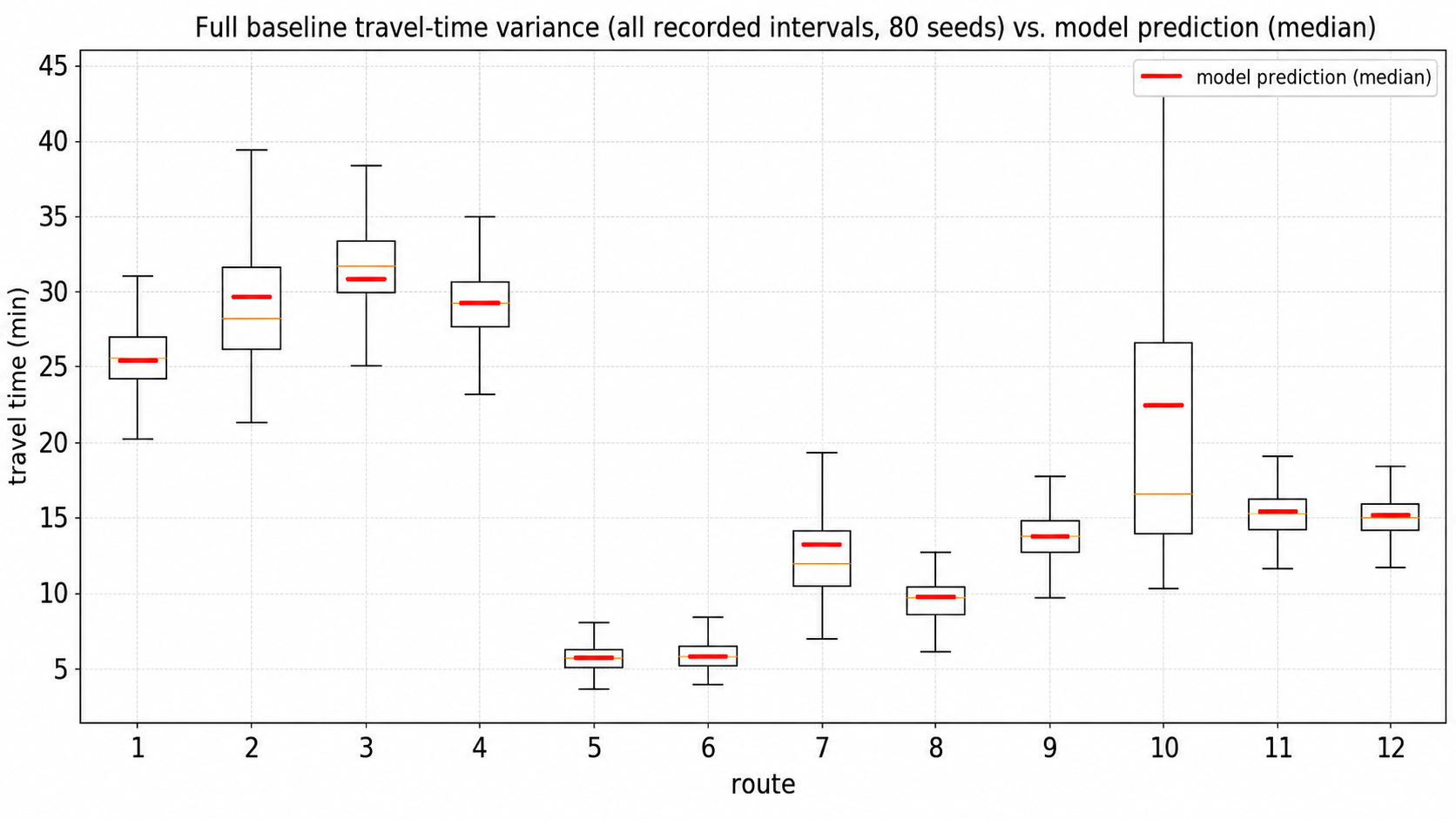}
  \caption{Baseline travel-time distribution across 80 seeds (box plots) versus Model 1's median prediction (red) for each of the twelve monitored routes.}
  \Description{Box plots of baseline travel time per route across 80 simulation seeds, with the model's median prediction overlaid as a red marker.}
  \label{fig:baseline_variance}
\end{figure}

Because travel times vary across simulation seeds even under identical baseline conditions, MAE alone does not indicate whether prediction error exceeds the network's inherent variability. Figure~\ref{fig:baseline_variance} compares the distribution of travel times across all 80 baseline simulations against the model's median prediction for each route. For eleven routes, predictions fell within or near the interquartile range of the baseline distribution, indicating that model error was comparable to the network's natural variability under incident-free conditions. Route 10 again stood apart, exhibiting both the largest prediction error and the widest travel-time distribution, ranging from approximately 12 to over 40 minutes across seeds. 

Its distribution was strongly right-skewed, with a median near 15 to 16 minutes and a small number of extreme delays inflating the mean. The model's prediction fell above the route's median but below its outlier-inflated mean, consistent with the underprediction in Table~\ref{tab:baseline_routes}. This suggests Route 10's higher error reflects its unusually skewed travel-time distribution rather than a systematic weakness of the model.


\subsection{Model 2: Incident-Inclusive Performance}
\label{sec:model2results}

Evaluated on the full held-out test set (baseline and incident samples combined), the model achieved a relative MAE of 9.67\%. Model 1's 9.94\% is computed over baseline samples only, so the two figures are not a controlled comparison of the effect of incident training.

Restricting evaluation to incident-affected test samples specifically, the timesteps during which an active blockage was present, produced a relative MAE of 6.45\%. The twelve per-route relative MAEs behind that aggregate have an unweighted mean of 7.61\% and a median of 5.60\%; the gap reflects that the aggregate weights each route by its travel time while the unweighted mean does not. In absolute terms, MAE on the route affected by each blockage averages 119.9 seconds, or 2.0 minutes, across the twelve incident locations, ranging from 0.7 to 3.7 minutes (bold cells, Table~\ref{tab:mae_by_edge}). Absolute and relative figures are both reported because neither is sufficient alone: route travel times span 5.9 to 31.0 minutes, so a two-minute error represents 6\% on the longest route and 33\% on the shortest. Incident timesteps give the lower relative MAE, 6.45\% against 9.67\% over the full test set. This does not mean forecasts are more accurate during incidents. The error in seconds is no smaller; the percentage falls because trips take longer while a blockage is active, and the same error is a smaller share of a longer trip. Model 1 was not evaluated on incident samples, so no improvement over incident-free training is claimed.

Table~\ref{tab:mae_by_edge} reports MAE by disrupted road segment and evaluated route, with the route or routes directly affected by each disrupted segment highlighted; small differences between its incident-free reference row and the actual travel times in Table~\ref{tab:baseline_routes} reflect different subsets of the 80 baseline simulations underlying the two evaluations, consistent with the seed-to-seed variability already observed for these routes.

\begin{figure}[tb]
  \centering
  \includegraphics[width=\columnwidth]{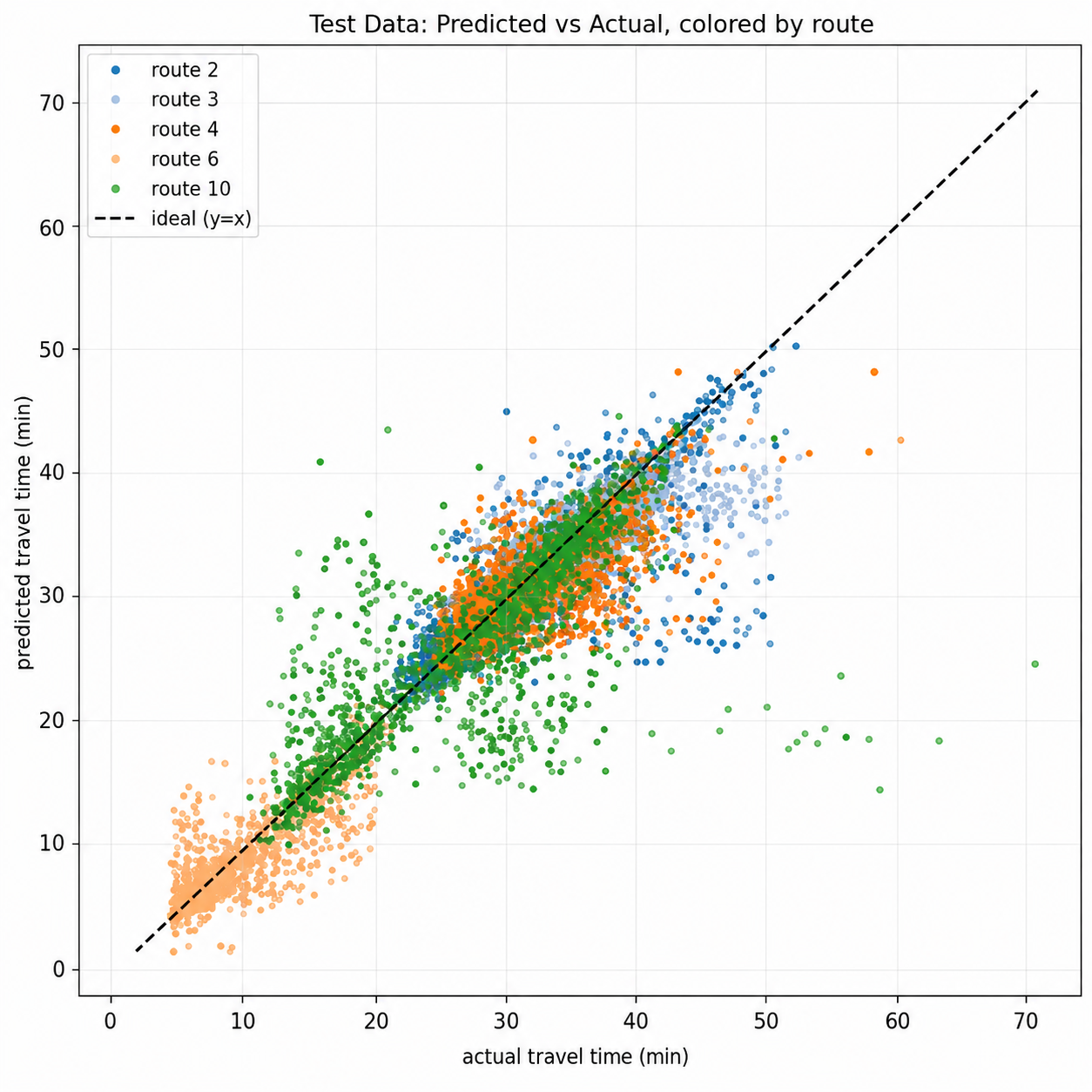}
  \caption{Model 2 predicted versus actual travel time across the full test set, colored by route, with the ideal $y=x$ line shown for reference.}
  \Description{Scatter plot of predicted versus actual travel time for the incident-inclusive model, colored by route.}
  \label{fig:pred_vs_actual}
\end{figure}

Figure~\ref{fig:pred_vs_actual} shows predicted versus actual travel time across the full test set, colored by route, illustrating that prediction quality is broadly consistent across routes of very different typical travel time, with routes 5 and 6, the shortest routes, showing the tightest clustering around the ideal line and route 10 showing the widest spread, consistent with the variance already observed in Model 1.

\begin{table*}[t]
  \centering
  \Description{Table of mean absolute error in seconds for each of twelve routes under twelve different disrupted-edge scenarios, with directly affected routes highlighted.}
  \label{tab:mae_by_edge}
  \begin{tabular}{lcccccccccccc}
    \toprule
    Edge & R1 & R2 & R3 & R4 & R5 & R6 & R7 & R8 & R9 & R10 & R11 & R12 \\
    \midrule
    -117879      & 99.66  & \textbf{66.96}  & 165.76 & 92.62  & 48.08 & 102.36 & 96.40  & 42.67 & 75.94 & 326.42 & 62.04 & 48.39 \\
    -117959      & 109.70 & \textbf{115.57} & 168.28 & 102.55 & 32.71 & 102.96 & 91.85  & 46.00 & 83.61 & 317.33 & 62.66 & 47.09 \\
    -124860      & 94.37  & \textbf{119.87} & 178.39 & 97.82  & 30.82 & 108.19 & 90.73  & 42.81 & 80.41 & 325.95 & 69.25 & 51.48 \\
    -119869      & 92.35  & 121.98 & \textbf{109.93} & 147.69 & 79.05 & 100.75 & 100.74 & 43.45 & 81.24 & 291.15 & 64.54 & 43.45 \\
    -B14.136     & 93.47  & 123.37 & \textbf{153.87} & 146.26 & 31.72 & 107.64 & 88.13  & 40.55 & 82.54 & 321.52 & 77.07 & 59.50 \\
    -112343      & 94.90  & 129.41 & 164.21 & \textbf{91.27}  & 46.58 & 107.43 & 97.42  & 41.08 & 79.82 & 299.09 & 74.34 & 61.44 \\
    -123926      & 98.70  & 133.35 & 156.98 & \textbf{73.58}  & 39.72 & 111.41 & 103.12 & 45.87 & 79.77 & 328.80 & 62.50 & 46.15 \\
    -B64.41      & 89.57  & 133.86 & 170.24 & \textbf{188.29} & 29.99 & 94.39  & 97.41  & 46.60 & 79.08 & 312.97 & 56.78 & 48.50 \\
    -125436\_net2 & 98.79 & 125.29 & 154.62 & 101.59 & 98.81 & \textbf{43.09} & 96.27  & 47.30 & 77.47 & 307.32 & 63.36 & 50.36 \\
    -114944      & 91.60  & 116.46 & 147.78 & 93.24  & 34.78 & \textbf{120.25} & 88.50  & 46.47 & 76.96 & 371.82 & 65.56 & 49.19 \\
    -117078      & 95.95  & 131.57 & 162.08 & 105.65 & 35.59 & 103.02 & 101.02 & 46.19 & 88.06 & \textbf{221.34} & 66.58 & 47.62 \\
    -122441      & 97.43  & 135.20 & 156.09 & 107.86 & 51.53 & 110.10 & 94.39  & 45.46 & 89.13 & \textbf{134.24} & 63.76 & 44.85 \\
    \midrule
    Baseline (s) & 1550.08 & 1857.60 & 1828.17 & 1730.62 & 356.76 & 366.17 & 793.47 & 574.97 & 826.61 & 1409.72 & 908.46 & 895.89 \\
    \bottomrule
  \end{tabular}
\end{table*}


\subsubsection{Forecast Behavior During Individual Incidents}

\begin{figure*}[t]
  \centering
  \includegraphics[width=\textwidth]{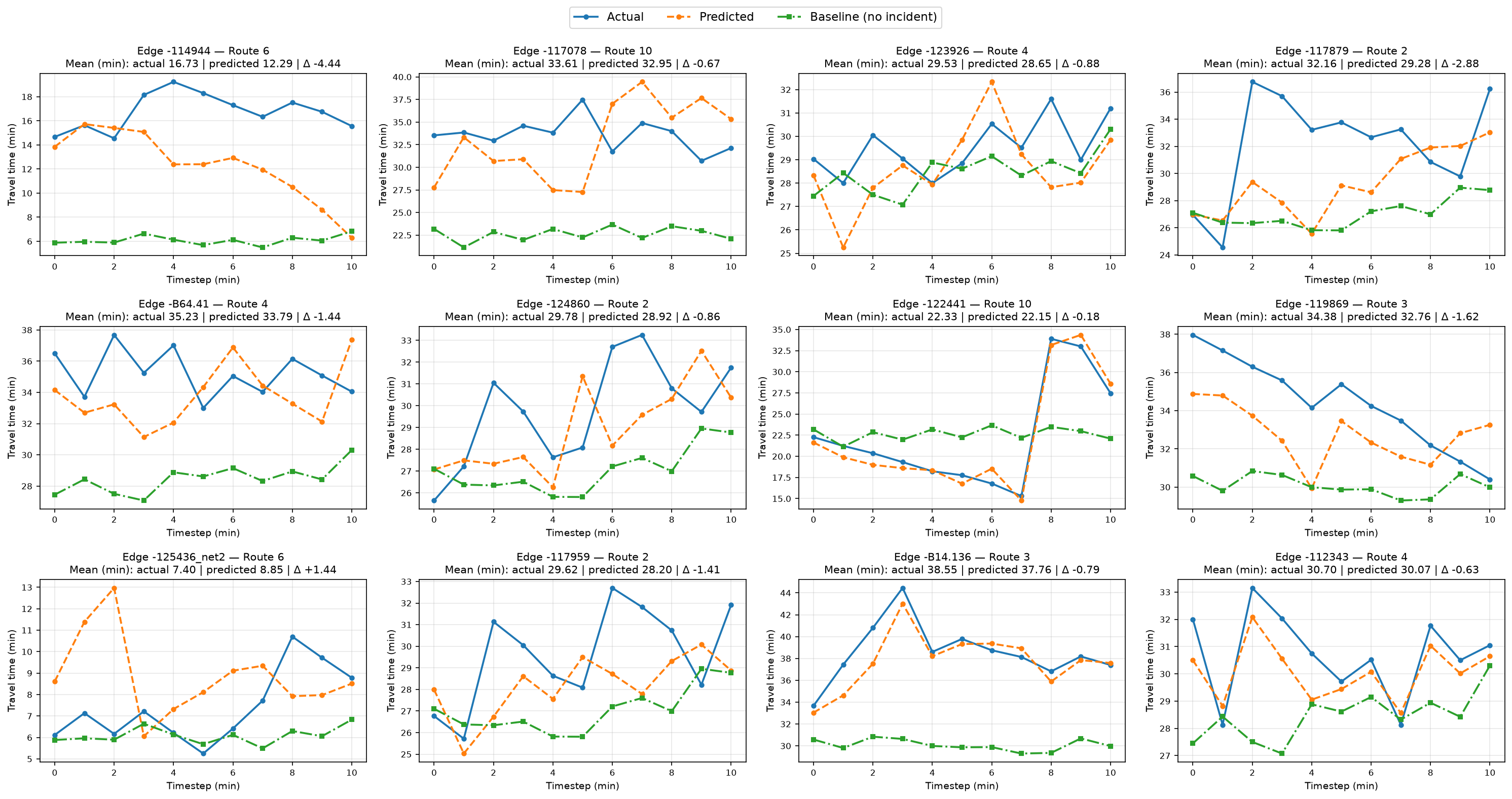}
    \caption{Model 2 forecasts during active incidents: actual travel time, the model's prediction, and the incident-free baseline, one panel per training incident location at 40-minute duration and a single seed. Routes match the bold cells in Table~\ref{tab:mae_by_edge}; y-axis ranges are per panel.}
  \Description{Grid of twelve line plots, one per disrupted road segment, each showing actual travel time, predicted travel time, and the incident-free baseline over an eleven-minute forecast horizon during an active forty-minute lane blockage.}
  \label{fig:model2_curves}
\end{figure*}

Aggregate error statistics describe average accuracy but not how the model behaves during a specific disruption, which is the behavior that matters for the applications described in Section~\ref{sec:background}. Figure~\ref{fig:model2_curves} therefore shows one representative 40-minute blockage at each of the twelve training incident locations. For each, the input window is the first 15-minute observation window falling entirely inside the active blockage period, and the plotted horizon is the 11 minutes the model predicts from it, that is, minutes 5 through 15 after that window ends. Alongside actual and predicted travel time, each panel shows the incident-free baseline for the same route at the same clock minutes, averaged over all 80 baseline seeds, so that the elevation caused by the incident is visible rather than implied.

Two behaviors are apparent. First, the model reproduces incident-induced elevation rather than reverting to typical conditions: in every panel the predicted curve sits well above the incident-free baseline, and for the most severely affected cases, such as segment -117078 on route 10, where actual travel time reaches roughly 33 minutes against a baseline near 23, the prediction tracks the elevated level rather than the baseline. Second, accuracy is uneven in a way that aggregate statistics conceal. The figures in the remainder of this paragraph are relative differences between mean predicted and mean actual travel time over the plotted horizon, not relative MAEs, since averaging before differencing allows errors to cancel. Mean predicted travel time falls within 5\% of mean actual for nine of the twelve locations and within 3\% for six of them, while the two largest relative differences, $-26.6\%$ for segment -114944 and $+19.5\%$ for segment -125436\_net2, both occur on route 6. Route 6 has the second-shortest baseline travel time in the network at approximately six minutes, so a fixed absolute error translates into a large percentage there: the absolute deviations in these two cases, 266 and 87 seconds, are comparable to those on routes several times longer that report differences below 5\%. This is the same effect that makes relative MAE a route-composition-sensitive summary, and it is worth separating from genuine differences in model skill.

The panel for segment -124860 illustrates a case worth stating explicitly, because it also explains that segment's attribution in Table~\ref{tab:mae_by_edge}. This blocked segment is not itself traversed by route 2. It terminates at a junction that route 2 passes through, so the disruption reaches the route as cross-street queuing at a shared junction rather than as a blockage along the route itself. The model predicts the resulting elevation to within 2.9\% in mean travel time, indicating that the learned response is not restricted to blockages lying on the monitored path, which is the behavior a network-wide screening application depends on.

Panel-to-panel variability in the actual curves is also visible and is expected: each panel is a single simulation seed rather than an average over seeds, so it retains the stochastic variation that Figure~\ref{fig:baseline_variance} quantifies for baseline conditions.

\subsubsection{Performance on Previously Unseen Locations}
\label{sec:transfer}

\begin{figure*}[t]
  \centering
  \includegraphics[width=\textwidth]{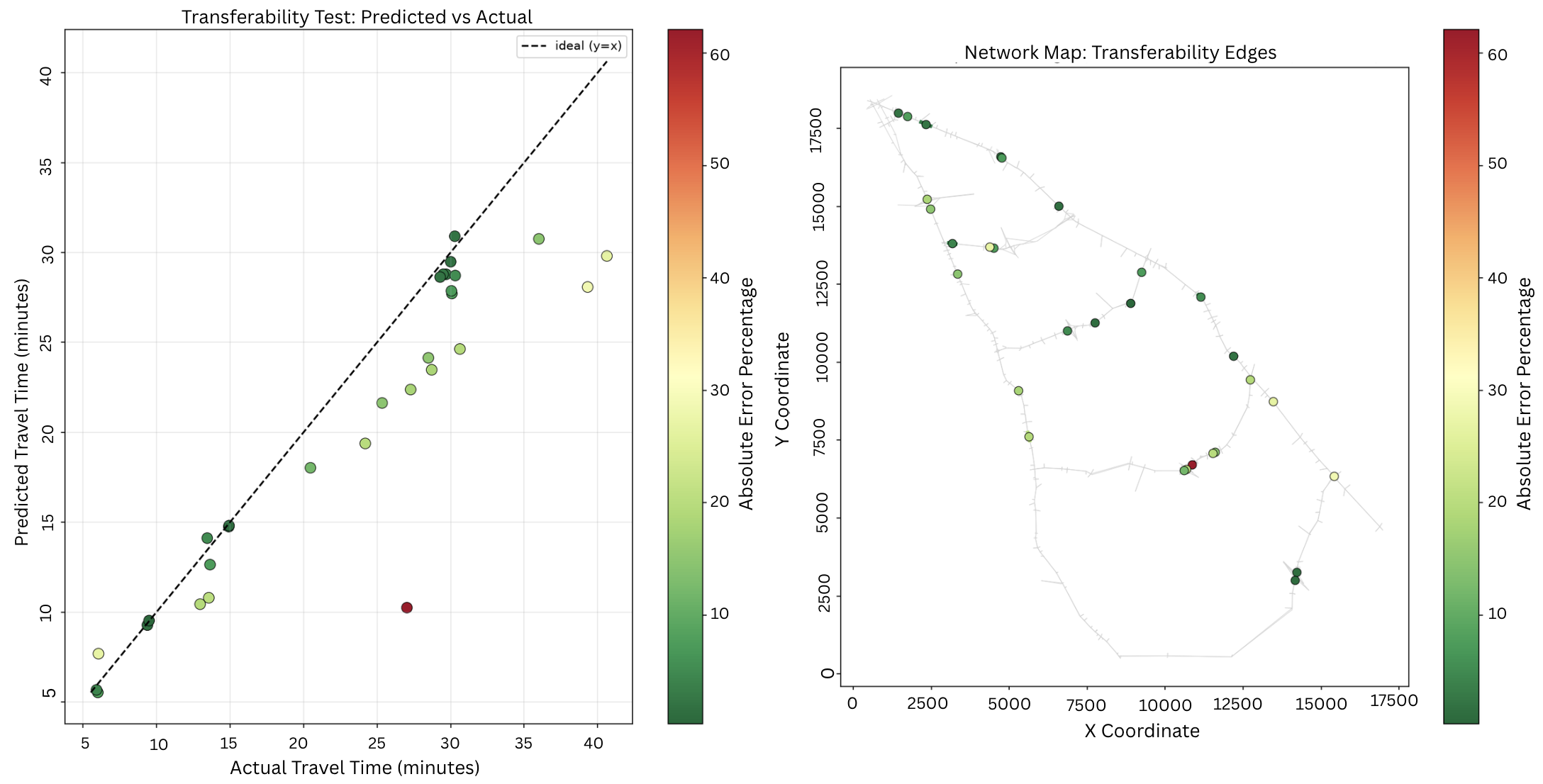}
  \caption{Left: predicted versus actual travel time for the thirty unseen-location test segments, colored by absolute error percentage. Right: the same thirty segments shown in their network position, colored identically, showing no strong spatial clustering of high-error locations.}
  \Description{Two-panel figure. Left panel is a scatter plot of predicted versus actual travel time for unseen edges, colored by error. Right panel shows the same edges positioned on the Nashville road network map, colored by the same error scale.}
  \label{fig:transfer_map}
\end{figure*}

To evaluate whether learned incident response generalizes beyond the locations used in training, the model was additionally tested on thirty incident locations excluded entirely from training, at a fixed 40-minute duration. This evaluation achieved a relative MAE of 13.62\%, with most per-segment relative MAEs falling between 7 and 9\%. A smaller subset of segments was substantially worse, including one outlier location at 63.94\% (Figure~\ref{fig:transfer_map}), giving a pattern of strong performance at most unseen locations alongside a specific subset of harder cases.

Generalizing beyond the locations represented in training is a recognized limitation of this model class. Most spatio-temporal graph models are transductive, assuming a fixed sensor configuration, and recent work addresses the limitation directly by designing inductive predictors that forecast at locations never observed during training~\cite{zhou2025mogernn, roth2022funs}. A direct numerical comparison with those results would not be meaningful here, for two reasons. First, reported degradation depends heavily on evaluation protocol, particularly whether test data are drawn from time periods also seen during training, so figures are not comparable across studies unless the protocols match. Second, the transfer evaluated here is to unseen \emph{incident} locations within a network the model has otherwise seen, which is a different and less demanding form of generalization than forecasting at sensor locations with no observation history. The results are therefore reported on their own terms rather than benchmarked against published ranges.

\section{Discussion and Conclusion}

Two STGCN models were trained to forecast route-level travel times on a simulated Nashville, Tennessee network using directional counts at its 129 signalized intersections. The baseline model reached a test MAE of 109.68 seconds, a relative MAE of 9.94\%. The incident-inclusive model reached 9.67\% relative MAE across its combined test set and 6.45\% on active-incident timesteps.


Two error patterns should be separated from genuine modeling weakness. Route 10 dominates the baseline error, but its travel-time distribution is right-skewed with a few extreme delays, and the prediction falls between its median and its outlier-inflated mean. Relative MAE is also worst on the shortest routes, where a small absolute error becomes a large percentage. This matters little for re-routing, where absolute error governs, but it makes relative MAE sensitive to route composition.


Transfer to thirty withheld incident locations gave 13.62\% relative MAE aggregated over all thirty, though the typical segment fell between 7 and 9\%. The gap between the two is a minority of badly predicted locations, one reaching 63.94\%. The network-wide incident flag, rather than one localized to the disrupted segment, requires the model to infer where effects are strongest from local traffic alone, which may explain the spread. Identifying what distinguishes the harder locations is the direct next step, and is where inductive formulations for unobserved locations~\cite{zhou2025mogernn, roth2022funs} become relevant. A broader limitation bounds every result reported here: training and evaluation are entirely within simulation. Input was restricted to quantities measurable at signalized intersections so that this gap can be closed, but showing that accuracy survives real sensor noise, dropout, and detector faults remains future work~\cite{saroj2023imputation}. Both contributions claimed at the outset are borne out: route travel time is forecast from signalized-intersection counts alone, at 9.94\% relative MAE under ordinary conditions, and incident response transfers to thirty locations never seen in training at 13.62\%, with the harder locations identified rather than averaged away. Because the model consumes only data agencies already collect and returns a forecast in tens of milliseconds, it is usable as a fast surrogate for resilience screening, incident management, and proactive re-routing, none of which are practical with microscopic simulation alone. 
Whether a single model genuinely serves both regimes remains open: settling it requires evaluating each model on the other's test population, which would show whether incident training costs ordinary-traffic accuracy or buys anything during disruption. Work is ongoing on a third model trained across multiple demand levels, on training with the full incident dataset, and on replacing the network-wide incident flag with a spatially localized signal.

\begin{acks}
This work was supported at Oak Ridge National Laboratory in part by the  DOE Office of Science, Office of Workforce Development for Teachers and Scientists, through the Science Undergraduate Laboratory Internships Program and in part by the DOE, Office of Critical Minerals and Energy Innovation, Transportation Technologies Office. The authors thank Guanhao Xu, Jinghui Yuan, Wan Li, and Ross Wang for the Nashville SUMO simulation model and Nashville Department of Transportation for traffic data.

\end{acks}

\bibliographystyle{ACM-Reference-Format}
\bibliography{references}

\end{document}